\documentclass[doublecol]{epl2} 

\usepackage{amsmath}
\usepackage{amssymb}
\usepackage{epsfig}
\usepackage{graphics}
\usepackage{graphicx}
\usepackage{paralist}
\usepackage{hyperref}
\usepackage[all]{hypcap}
\usepackage{xcolor}
\usepackage{tabularx}

\usepackage{dcolumn}   
\usepackage{bm}        

\title{Crossovers in supercooled solvation water: Effects of hydrophilic \& hydrophobic interactions}
\shorttitle{Crossovers in supercooled solvation water} 

\author{John Tatini Titantah\inst{1} \and Mikko Karttunen\inst{2}}
\shortauthor{Titantah \& Karttunen }

\institute{                    
  \inst{1}\!\!\!\!Department of Applied Mathematics, the University of Western Ontario, 
                 1151 Richmond Street North, London, Ontario N6A\,3K7, Canada\,\,
  \inst{2}\!\!\!\!Department of Chemistry \& Waterloo Institute for Nanotechnology, 
                  University of Waterloo, 200 University Avenue West, Waterloo, Ontario, Canada N2L\,3G1
}

\abstract{
Systematic 8\,ns \textit{ab initio} molecular dynamics (AIMD) were performed to study 
the structure and dynamics of water in bulk and close to
hydrophobic (CH$_\mathrm{3}$) and hydrophilic (carbonyl) groups 
of tetramethylurea (TMU). 
Dynamical behaviour showed two crossovers: The first around the
hydrophobic group  at $T_X\! =256 \! \pm 4$\,K, and the second at $265\pm$5\,K 
related to the relative strengths of water-water and water-carbonyl hydrogen bonds (HBs).
For bulk water, relaxation times appear to diverge
at $T_c\! =\! 213\!\pm \!10$\,K, rendering support to the
liquid-liquid critical point hypothesis.
To identify the effects due to the hydrophilic carbonyl group, systems of water
with one methane molecule were used as references.
Our findings are related to the structural and thermodynamic transitions 
reported for proteins in solution and may play a role in 
cold denaturation.
}

\begin{document}

\maketitle

\section{Introduction}

Although possibly the most investigated substance, water still holds many 
puzzles.  For bulk water, one of the fundamental questions is the proposed
existence of a liquid-liquid critical point at $T_c\approx 225$\,K~\cite{poole92}.
Proposed by Poole \textit{et al.} already in 1992~\cite{poole92}, it remains a topic of
intense interest~\cite{taschin13,mallamace13,Limmer2014}.
In addition to the fundamental properties of bulk water, 
precise information about the interplay between hydrophobic and hydrophilic interactions 
is crucial for understanding the physical mechanisms in biological
systems.  Proteins and lipids 
contain both hydrophobic 
and hydrophilic parts that respond to the presence of water differently depending on 
the thermodynamic and physiological conditions.
For example,  
Toppozini~\textit{et al.}  reported a series of melting transitions using 
neutron scattering with  cooling and heating of  lipid 
bilayers~\cite{toppozini12}: A melting transition at 252\,K for the acyl tail dynamics 
and solvation water, and at 264\,K for lipid diffusion. 
Mazza \textit{et al.} reported two distinct crossovers at
$\approx 252$ and 181\,K
in protein hydration~\cite{mazza11}.

Small hydrophobes are perhaps the simplest systems for investigating the physical origins of such effects.
In elegant experiments, Qvist and Halle used  $^2$H nuclear magnetic resonance (NMR) to 
demonstrate that solvation water around small hydrophobes
undergoes a crossover from high activation energy 
to low activation energy rotation 
(with respect to bulk H$_2$O) at $T_X \!= \! 255 \! \pm \! 2$\,K~\cite{qvist08}.  
They predicted that far below $T_X$, solvation water should 
rotate faster than bulk water, thus at low temperatures contradicting the famous 
iceberg model of hydration\cite{Frank1945} 
which
has been demonstrated in higher temperatures by femtosecond mid-infrared (fs-IR) 
spectroscopy by Rezus and Bakker~\cite{rezus07}.
Qvist and Halle found $T_X$ to be close to the  dynamic  crossover temperature of  $\approx$252~\,K 
reported for protein solvation and attributed to fluctuations in HB formation~\cite{mazza11},  
and a melting peak reported for myoglobin between 250\,K and 260\,K~\cite{doster86}.  
Classical MD has been used extensively to explore
the low temperature behaviour of water
\cite{poole92,cuthbertson11,holten12,kumar07,russo14}.
The temperatures at which water's anomalies occur are, however, strongly force-field dependent: 
Commonly used water models put the melting point 
between 190 and 274\,K~\cite{vega05}.

 \begin{figure}
\begin{center}
\includegraphics
[width=6.7cm]
{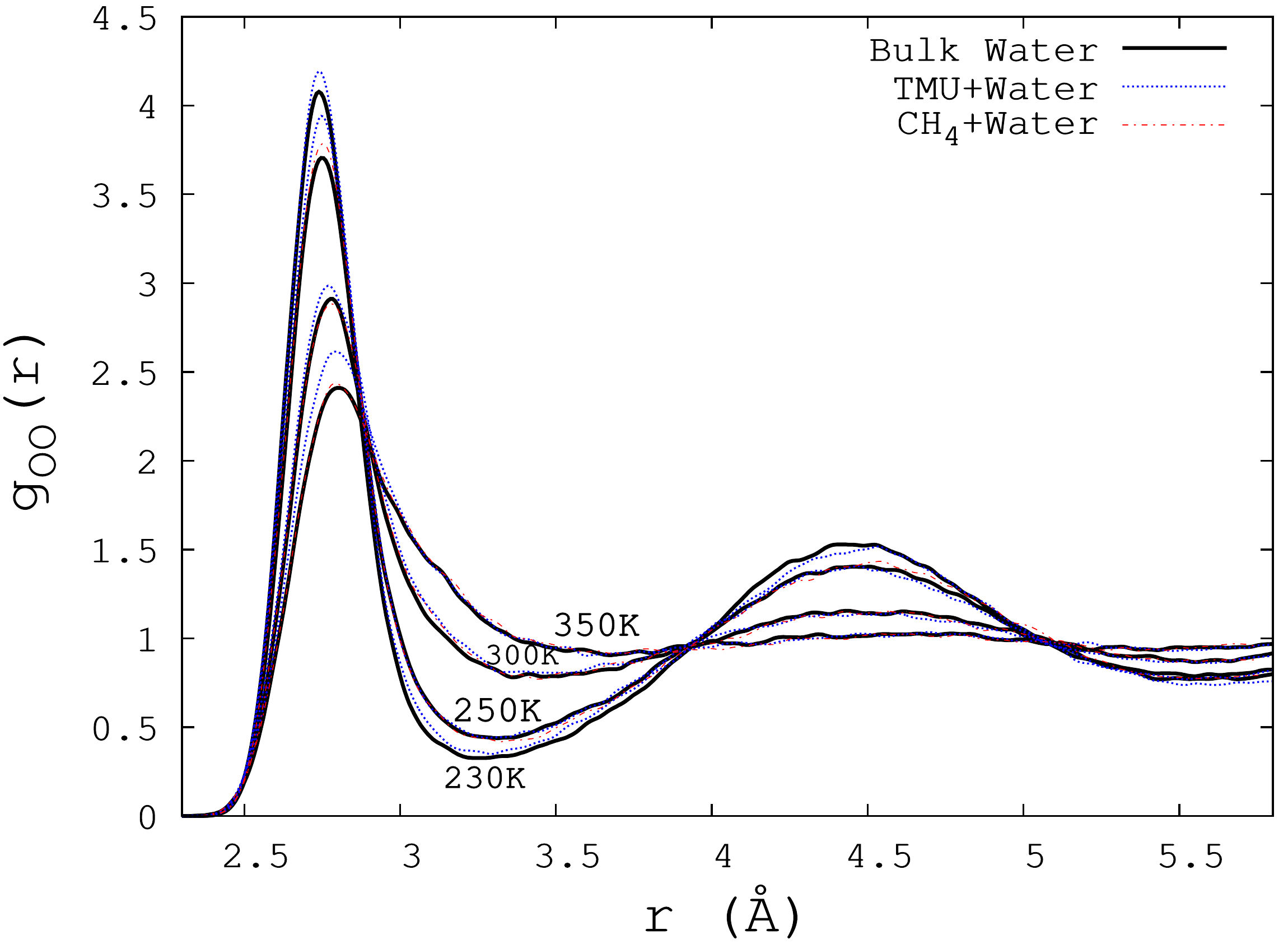} 
\vspace*{-0.2cm}
\caption{Water's oxygen-oxygen pair correlation in bulk water (at 230, 250, 300 and 350\,K), 
TMU+water system (at 230, 250, 300 \& 350\,K) and CH$_\mathrm{4}$+water system (at 250, 300 \& 350\,K).}
\label{fig:gr}
\end{center}
\end{figure}

\section{Simulation details}

We used AIMD  within the Born-Oppenheimer 
approximation to investigate the temperature dependence of 
dynamic and structural properties of water solvating a hydrophobic TMU 
for temperatures from 230 to 370\,K.
Crossover behaviours in bulk water, and water close to both hydrophobic
and hydrophilic groups were studied. The total simulation time was over 8\,ns
making this
among the most comprehensive AIMD studies of this kind. 
The CPMD code~\cite{cpmdcode} in the NVT ensemble was used.
The BLYP (Becke-Lee-Yang-Parr)~\cite{becke1988,lee1988} functional was used with
van der Waals interactions via the DFT-D3 parametrization~\cite{hujo13}. For C, O and N, Troullier-Martins~\cite{troullier91}
pseudopotential was applied, and Kleinman-Bylander~\cite{kleinman82} was used for hydrogen. 
Time step of 0.121\,fs was used and the production simulations 
ranged from more than 100\,ps at high temperatures to over 500\,ps at low 
temperatures.  The production runs were preceded by conjugate 
gradient relaxation of positions and wave functions, 
and at least 50\,ps equilibration. Other simulation details are as in refs.~\cite{titantah2012,titantah2013}.
Reproducibility is illustrated 
in fig.~\ref{fig:gr} where  O-O pair correlations for water molecules in bulk, 
in the water+TMU system, 
and in the CH$_\mathrm{4}$+water system 
are shown. Beyond the 1$^\mathrm{st}$ coordination shell, all the correlations for the same temperatures are practically identical. 
The difference in the 1$^\mathrm{st}$  shell is a consequence of solute-induced 
translational ordering in the surrounding water.

\begin{figure}
\begin{center}
\includegraphics
[width=7.5cm]
{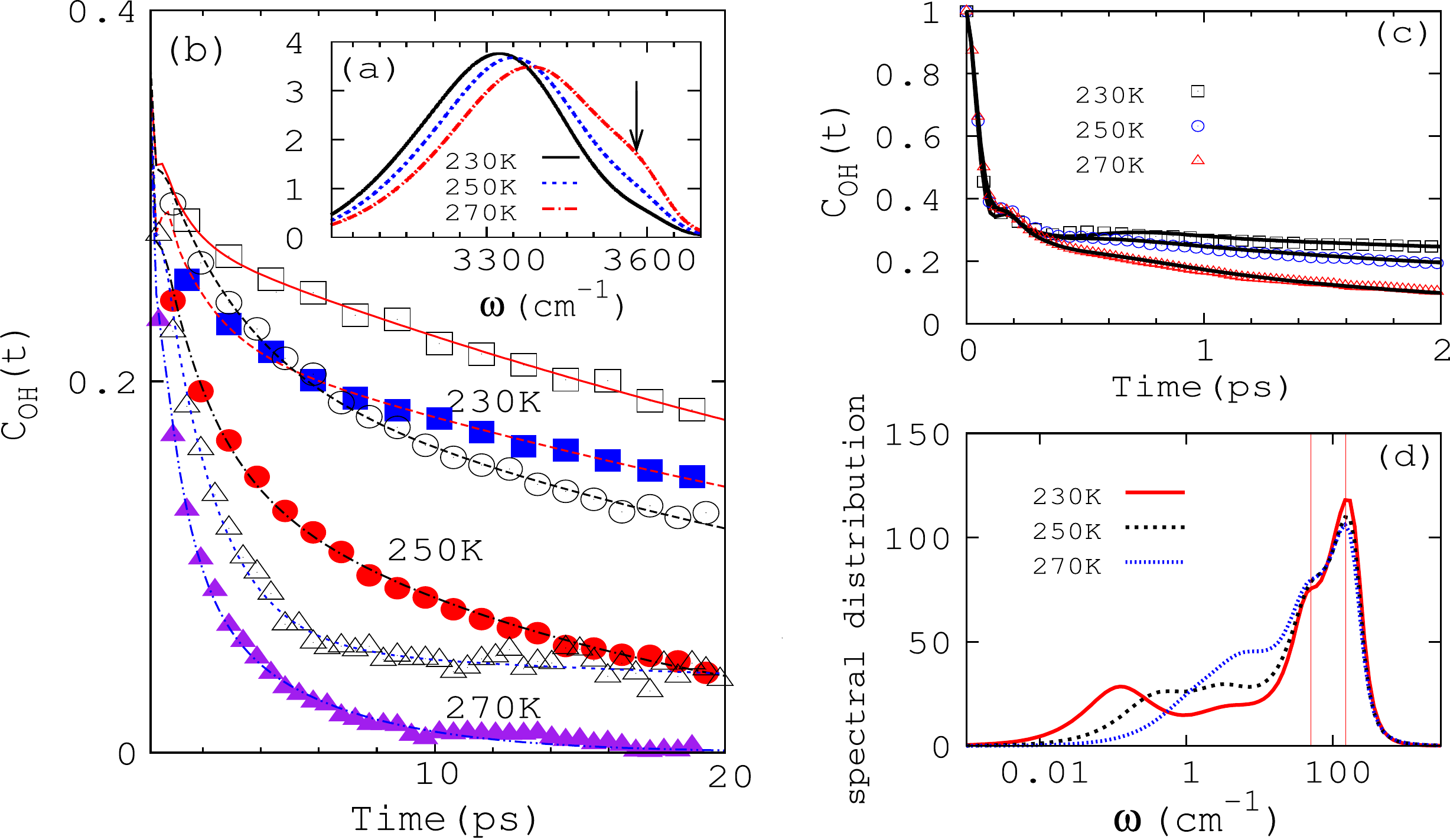} 
\vspace*{-0.2cm}
\caption{OH stretch frequency a) distribution, and 
b) time-correlation function for bulk water for $t>$0.15\,ps (filled symbols) and 
CH$_\textmd{3}$ contact water [$<4$~\AA~from the C atom of 
TMU CH$_\textmd{3}$ groups] 
(open symbols).
c) Short time behaviour of the bulk water correlation function, and (d) the spectral 
distribution obtained by Fourier transforming eq.~(\ref{eq:oh-corr}) 
for solvation shell water of TMU. 
Lines in b) \& c): Fits with eq.~(\ref{eq:oh-corr}). 
Vertical lines in d): 50 cm$^{-1}$ \& 150 cm$^{-1}$.}
\label{fig:oh-corr}
\end{center}
\end{figure}

\section{ OH stretch frequency}

We start from the time correlations of OH stretch frequency vibrations ($\omega(t)$):
$C_\mathrm{OH}(t)\!\!=\!\!\left<\Delta\omega(t)\Delta\omega(0)\right>\!/\!\left<\Delta\omega(0)^2\right>$, 
with $\Delta\omega(t)\!=\!\omega(t)\!-\!\left<\omega\right>$, and
$\left<\omega\right>$ is the 
average over time and OH groups. $C_\mathrm{OH}(t)$ is accessible by fs-IR~\cite{fecko03,fecko05}.  
We performed a  time series analysis, as introduced by  Mallik \textit{et al.}~\cite{mallik08}, 
to determine the time-dependent OH stretch vibrational frequencies.
Since this method underestimates the OH stretch frequency  of water by about 
6~\%~\cite{mallik08,titantah2013}, the frequencies were rescaled by a factor of 1.06.

Figure~\ref{fig:oh-corr}a shows the time-averaged frequency distributions for bulk water.
As temperature decreases, water molecules vibrate with lower frequency, a signature of HB strengthening. 
Another interesting feature is 
the appearance and increase in the proportion of dangling OH bonds at $\approx$3600 cm$^{-1}$  
(arrow, fig.~\ref{fig:oh-corr}a) as temperature increases.

\begin{figure}
\begin{center}
\includegraphics[width=7.5cm]
{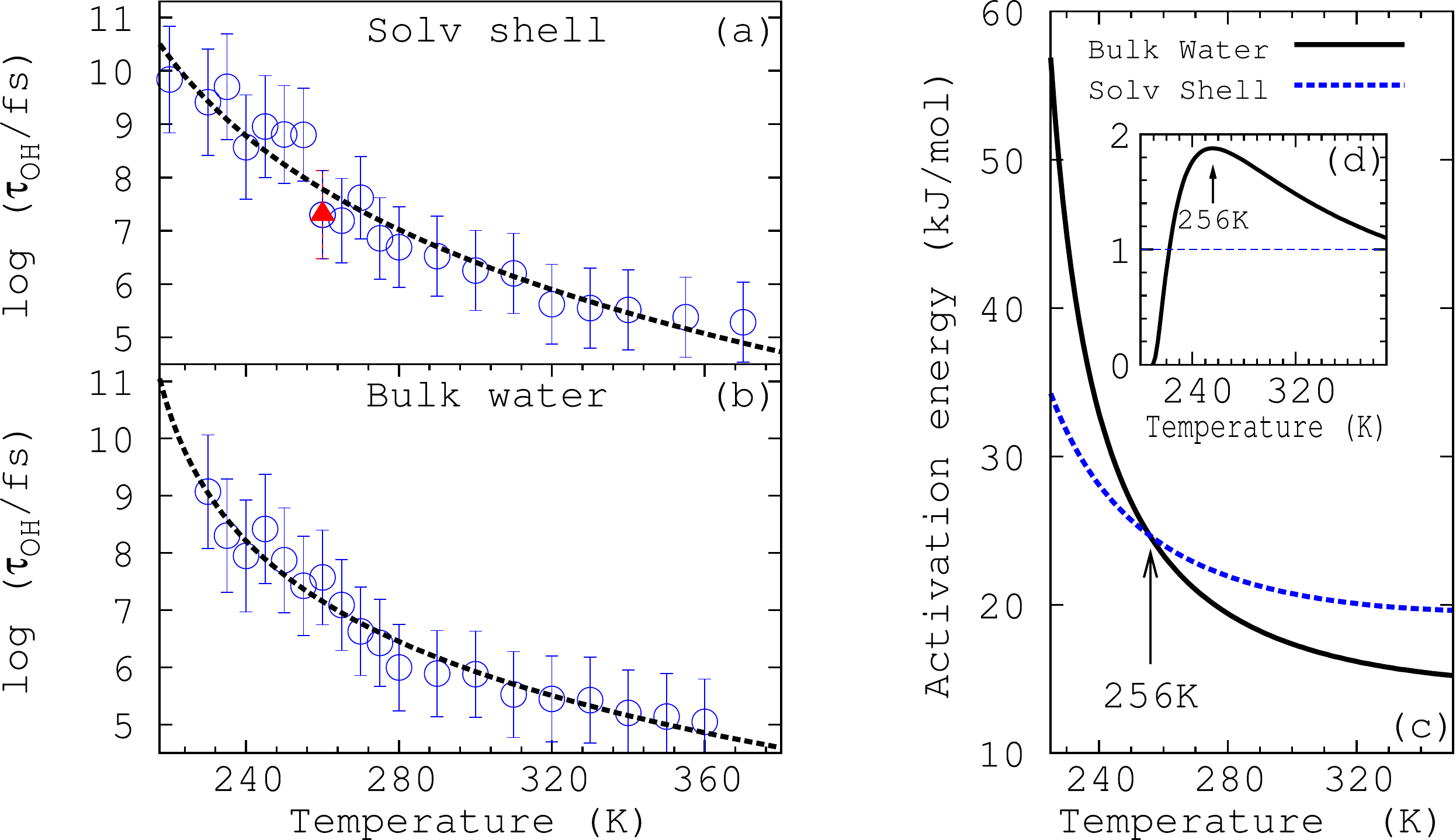} 
\vspace*{-0.2cm}
\caption{OH stretch frequency correlation time for 
a) water in contact with the TMU CH$_\textmd{3}$ groups ($<4~\mathrm{\AA}$~from the C atom of CH$_\textmd{3}$ group).
The red triangle at 260\,K: Independent control simulation. 
b) Bulk water. 
c) OH stretch vibrational activation energy showing a crossover at 256\,K.  
{Inset d): The perturbation ratio; 
maximum of 1.9 at 256\,K.} Lines 
in a) \& b): Fits using $\tau=\tau_0\left(T/T_c-1\right)^{-\gamma}$. 
Solvation water: $\tau_0=110\pm 20$ fs, $T_c=190 \pm 20$\,K and $\gamma=3.0 \pm 0.8$. 
Bulk water:  $\tau_0=68 \pm 10$ fs, $T_c=213 \pm 10$\,K and $\gamma=1.9 \pm 0.3$.  }
\label{fig:oh-time}
\end{center}
\end{figure}

\begin{figure}
\begin{center}
\includegraphics[width=6.8cm]
{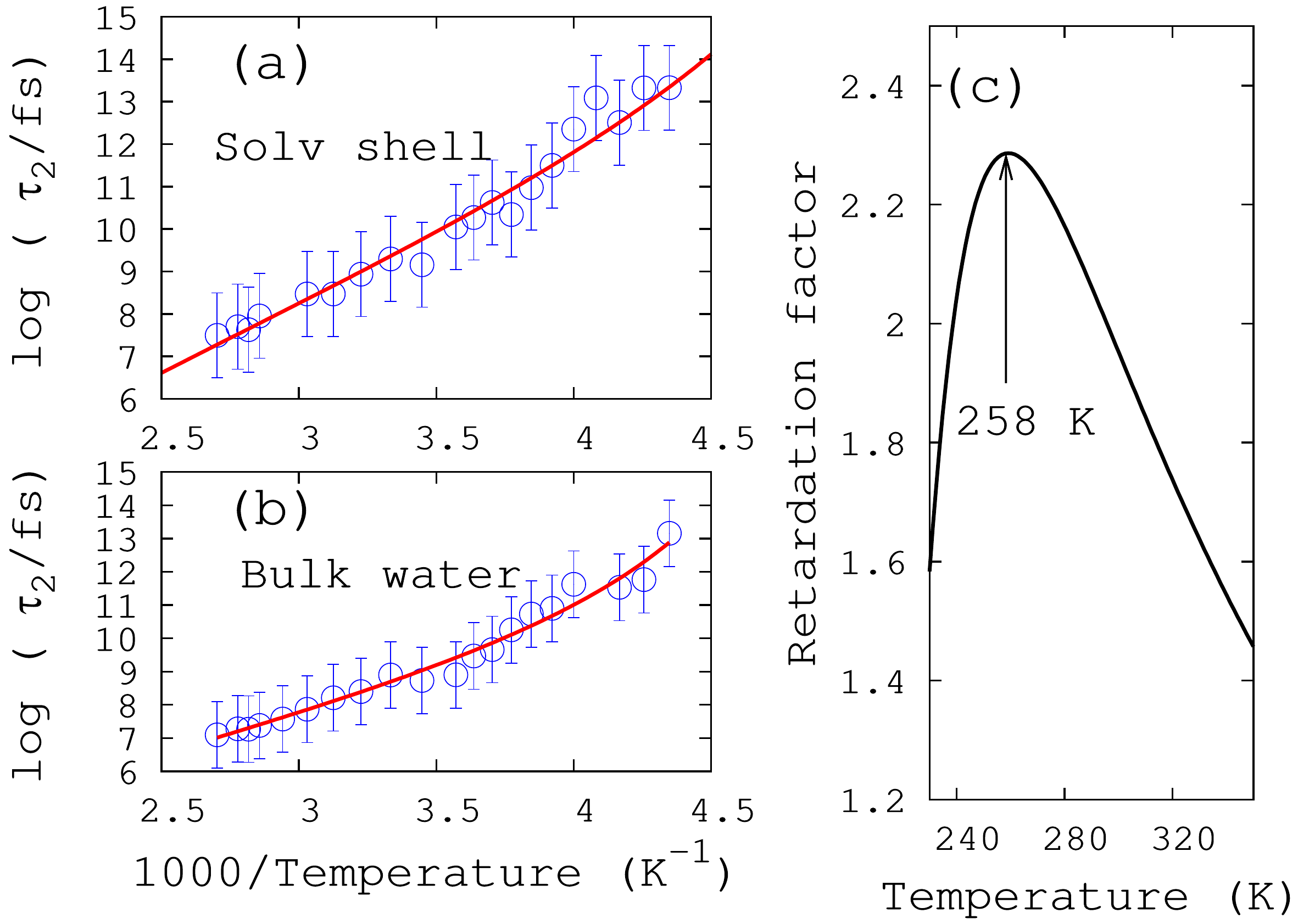} 
\vspace*{-0.2cm}
\caption{
The orientational correlation time $\tau_\mathrm{2}$ 
(from $C_2(t)$, see text) for water molecules 
in a) the solvation shell of TMU  and b) bulk water. 
c) The retardation factor 
shows a crossover at $T=258 \pm$5\,K. 
The lines in a) \& b) are fits to 
$\tau_\mathrm{2}(T)\!=\!\tau_0(T/T_\mathrm{c}\!-\!1)^{-\gamma}$. 
Bulk water:  $\tau_0\!\!=\!\!540$\,fs, $T_\mathrm{c}\!=\!210 \!\pm \! 10$\,K, 
$\gamma$=3$\pm$0.4. Solvation water: $\tau_0$=1800 fs, $T_\mathrm{c}$=180$\pm$15 K and $\gamma$=4.5$\pm$1.
}
\label{fig:oh-time2}
\end{center}
\end{figure}

\begin{figure}
\begin{center}
\includegraphics
[width=6.5cm]
{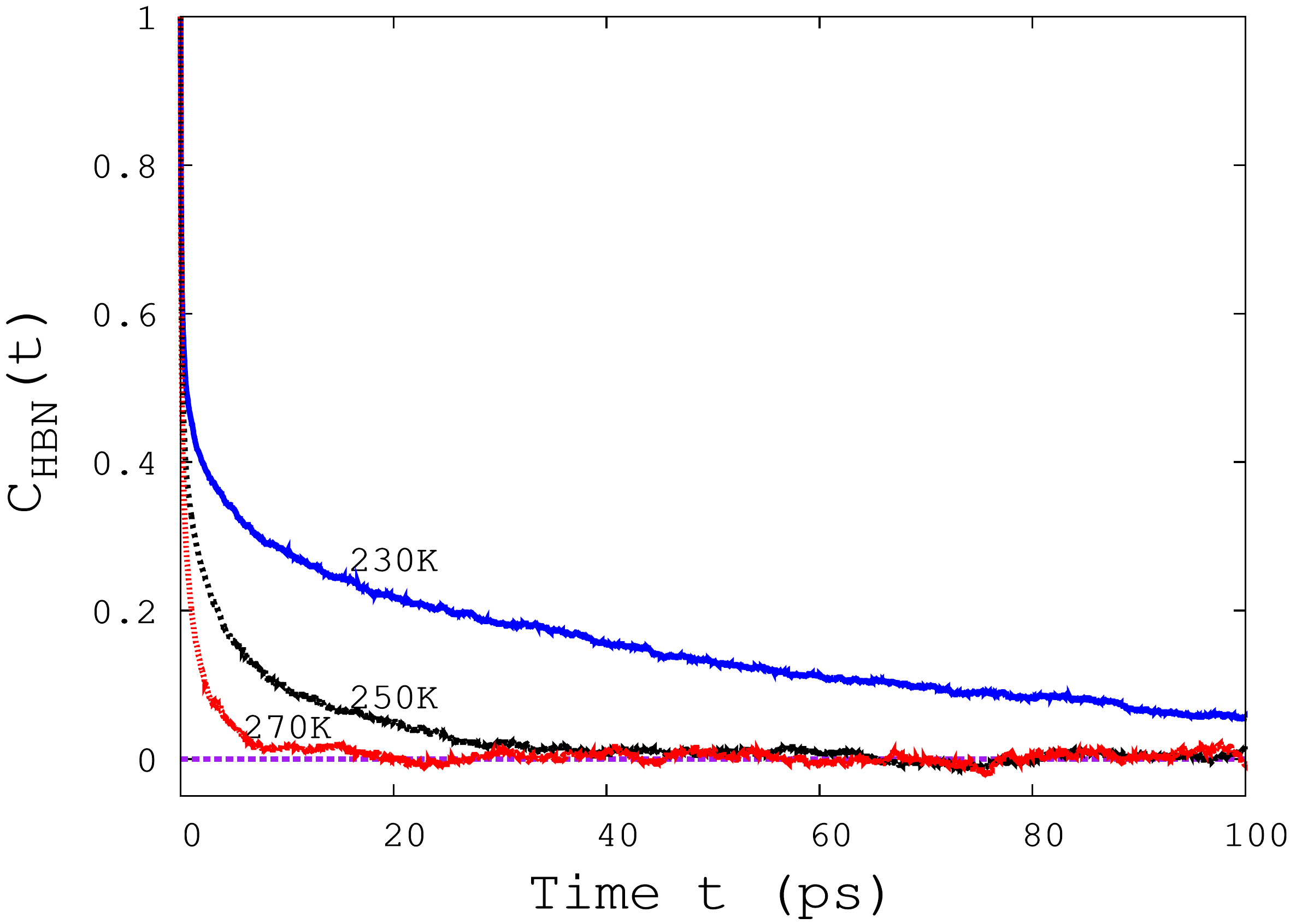} 
\vspace*{-0.2cm}
\caption{H-bond correlation for bulk water at 230, 250 and 270 \,K. At the 
lowest temperature of 230\,K HB dynamics is relaxed.}
\label{fig:hb-longcorr}
\end{center}
\end{figure}

\begin{table}
\resizebox{\columnwidth}{!}{%
\begin{tabular}
{|l|l|l|l|l|l|l|l|l|l|l|}
\hline
	$T$ & $a_0$ & $a_1$ & $a_2$ & $\lambda$ &  $\nu_\mathrm{HB}$ & $\nu_\mathrm{cage}$ & $t_0$ & $t_1$ & $t_2$ & $t_3$ \\
\hline
	230 & 0.304 & 0.326 & 0.086 & 3.7 & 143 & 49 & 0.09 & 0.13 & 0.57 & 42.20\\
\hline
	240 & 0.229 & 0.400 & 0.121 & 5.4 & 152 & 42 & 0.10 & 0.10 & 2.00 & 27.90\\
\hline
	250 & 0.133 & 0.466 & 0.139 & 4.9 & 153 & 46 & 0.12 & 0.09 & 1.11 & 20.50\\
\hline
	260 & 0.289 & 0.270 & 0.214 & 4.4 & 138 & 52 & 0.09 & 0.13 & 0.42 & 4.40\\
\hline
	280 & 0.223 & 0.162 & 0.299 & 5.3 & 155 & 62 & 0.10 & 0.14 & 0.16 & 1.92\\
\hline
	300 & 0.120 & 0.231 & 0.281 & 18.3& 134 & 61 & 0.12 & 0.10 & 0.15 & 1.44\\
\hline
	320 & 0.159 & 0.069 & 0.453 & 6.6 & 154 & 91 & 0.12 & 0.17 & 0.13 & 0.75\\
\hline
\end{tabular}}
\label{table:fit}
\vspace*{-0.2cm}
\caption{Fitting parameters for eq.~\ref{eq:oh-corr}. Temperature in Kelvins,
times $t_0-t_3$ in ps, and $\nu$'s in cm$^{-1}$. 
$a_0$ and $\lambda$ are dimensionless.}
\end{table}

$C_\mathrm{OH}(t)$ is shown in figs.~\ref{fig:oh-corr}b\,\&\,c.
For bulk, $C_\textmd{OH}$  
decays faster compared to solvation water, in agreement with anisotropy fs-IR measurements~\cite{rezus07,petersen09}.  
The correlation functions are characterized by weakly temperature-dependent  
fast 50-60\,fs initial decay and a  recoil at $\approx \! 175$\,fs,  plateauing between 200 and 500 fs 
and a long time decay tail beyond 500 fs.
Similar plateauing has been reported based on the 
mean-square displacement of water molecules for wait times of 300-500\,fs~\cite{gallo00}. 
Modeling the short-time dynamics with a damped oscillatory function has been suggested to explain spectral 
dynamics at 298\,K~\cite{moller04,mallik08}. 
To include the oscillations for the O-O-O caging effect, we modified the model to 
\begin{eqnarray}
C_\textmd{OH}(t)\!\!&\!\!\!\!\!\!=\!\!\!\!\!\!& \!\!
\left\{ a_0\exp\left(-{t/\tau_0}\right)
\left[\cos\left({2\pi t \nu_\mathrm{HB}}\right)
+\lambda\sin\left({2\pi t\nu_\mathrm{HB}}\right)\right]\right.\nonumber\\
&&+\left.\! a_1 \exp\left(-{t/\tau_1}\right)\right\}
\cos\left({2\pi t\nu_\mathrm{cage}}\right)\!+\!a_2\exp\left(\!-{t/ \tau_\beta}\right)\nonumber\\
& &+\left(1-a_0-a_1-a_2\right)\exp\left(-{t/\tau_\alpha}\right).
\label{eq:oh-corr}
\end{eqnarray}
Parameters are shown in Table~\ref{table:fit}.
The above equation reproduces the fast initial decay with time-scale $\tau_0$, the HB stretch recoil period $\nu_\textmd{HB}^{-1}$,  
and the plateau life-time $\tau_1$, a consequence of the cage vibrations (period $\nu_\textmd{cage}^{-1}$) of the 
surrounding water molecules. This latter is destabilized above 275\,K: 
This mode is more important in supercooled temperatures where more stable cages are preponderant. 
The parameter $\lambda$ 
accounts for the $\sim$15-20\,fs mismatch between echo-peak shift measurement and the OH  stretch-frequency 
correlation function~\cite{fecko05}. Cage dynamics is responsible for vibrations with wave number 
$k_\mathrm{cage}\negmedspace \sim \negmedspace 40\negmedspace -\negmedspace 60$\,cm$^{-1}$ 
usually found in the spectra of water~\cite{tsai05}. 
It is also very close to the recently reported 40 cm$^{-1}$ mode termed Boson peak 
in deep supercooled water~\cite{caponi09, kumar13};
{the Boson peak has also been observed in proteins, and associated with collective modes and
hydration-related multiple energy minima~\cite{joti05,kurkal06}}. 
The long time tail is modeled by two exponentials. 
The spectral functions 
are shown in fig.~\ref{fig:oh-corr}d.
For the spectral function, the Fourier transform of the time correlation was
multiplied by
$\omega \tanh(\hbar\omega/2/k_BT)$~\cite{Guillot1991}.
In deep supercooled temperatures (230 and 250\,K), four modes are found: Strongly temperature dependent $\alpha$ (0.01-1 cm$^{-1}$), 
the intermediate  $\beta$- (1-10 cm$^{-1}$), the cage mode (40-60 cm$^{-1}$), and 
the HB stretch mode (140-180 cm$^{-1}$)~\cite{tsai05}. 
The $\alpha$-mode is barely seen as a tail at 270\,K.
The mode at 0.01-1 cm$^{-1}$ compares well 
with depolarized light scattering measurements~\cite{comez13}.

Correlation times were obtained by integrating eq.~\ref{eq:oh-corr}. Their temperature dependencies 
are shown in figs.~\ref{fig:oh-time}a \& b for CH$_\textmd{3}$ contact water 
and bulk water, respectively.  To check reproducibility, independent  trajectories were generated at 260\,K.
Figure~\ref{fig:oh-time}a shows the two correlation times obtained from two independent trajectories. They are practically identical, 
showing the robustness of the results. 

The critical temperature of 213$\pm$10\,K for bulk water is less than  the reported  228\,K\cite{torre04},  
227\,K\cite{torre04,taschin13} \& 225\,K\cite{chu10}, although within margin of error. The large uncertainty 
is an inherent difficulty of the AIMD approach at low temperatures. 
Although not conclusively, the apparent divergence of dynamic properties lends support  to 
the liquid-liquid critical point scenario in deep supercooled water~\cite{poole92}.  

\section{Activation energies \& HBs}

Activation energies were determined using
$
E_\textmd{a}(T)\!=\!-k_\textmd{B}T^2{d\over d T} \log{ \tau(T)\over \tau_0}.
$
%
Results for bulk and contact water are shown 
in fig.~\ref{fig:oh-time}c. In agreement with Qvist and Halle~\cite{qvist08} and 
Tielrooij \textit{et al.}~\cite{tielrooij10,petersen09}, activation energies for contact water are 
higher than for bulk water at high temperatures. As temperature is lowered below the crossover 
temperature 256$\pm$4\,K, the activation energy of bulk water becomes larger than that of solvation water. 
This is in excellent agreement with the $^2$H NMR result  $T_X\!\!=\!\!255 \pm$2\,K~\cite{qvist08}.  
{As the inset fig.~\ref{fig:oh-time}d shows, the OH stretch correlation-based
perturbation factor $\tau_\mathrm{SolvShell}/\tau_\mathrm{Bulk}$ 
has a maximum of 1.9 at 256\,K. We also computed the retardation factor, that is, ratio of the 
solvation shell and bulk correlation times computed from rotational dynamics as is typically done
in experiments. 
Rotational dynamics was obtained from 
the correlation function $C_2(t)=\left<P_2\left(\cos \theta (t)\right)\right>_0$, 
where $P_2$ is the 2$^\mathrm{nd}$ order Legendre's polynomial, $\theta$ the angle 
an OH bond vector sweeps through in time $t$ and $\left<\dots\right>_0$ 
stands for averaging over water molecules and time. 
The correlation times and the retardation factor are shown in fig.~\ref{fig:oh-time2}. 
For the latter, we obtained 2.28.
This compares well 
with the NMR  value of 2.09~\cite{qvist08}. 
Classical MD on  trimethylamine N-oxide (TMAO)-water system found a slightly 
lower crossover temperature of $T_X \!\!\sim \!\!245$\,K 
and a retardation factor of $\approx$1.5~\cite{laage2014}.
The NMR  result 
is
1.92~\cite{qvist08}.
\begin{figure}[htb]
\begin{center}
\includegraphics
[width=7.2cm]
{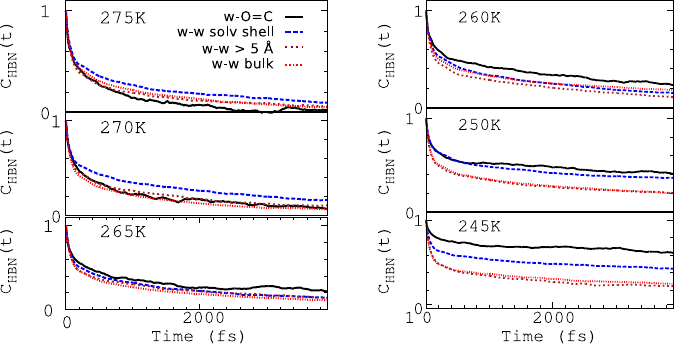} 
\vspace*{-0.2cm}
\caption{H$_2$O-H$_2$O and H$_2$O-carbonyl HB population correlation functions.
Water-water HBs: Water molecules 
closest to the CH$_\textmd{3}$ group ($<$4~{\AA} from CH$_\textmd{3}$), in bulk water, and 
for water molecules beyond 5~{\AA} from the CH$_\textmd{3}$ group. The water-carbonyl 
HBs become longer lived 
below 265\,K. Solvation water HBs are stronger than bulk water HBs 
and beyond 5~{\AA} from the CH$_\textmd{3}$ group, the water-water HB strength 
recovers the bulk water value indicating that hydrophobic effect is of short range.}
\label{fig:hb-corr}
\end{center}
\end{figure}

Crossovers have been reported in the properties of solvation water of proteins and 
lipids, and in the dynamics of lipid tails around $T_X$~\cite{mazza11,toppozini12,doster86}.
To study the effects of the hydrophilic carbonyl group, the HB lifetimes were examined. 
Three types of HBs were distinguished: \textit{(i)} between two water molecules in the bulk,  
\textit{(ii)} between water molecules in the solvation shell of the CH$_\textmd{3}$ group, and  
\textit{(iii)} between a water molecule and the oxygen atom of the carbonyl group. 
H-bonds were determined using the usual geometric criterion (distance less than the
first minimum of $g(r)$ and angle $<30^{\circ}$)~\cite{titantah2012}.
The time correlation of each HB population was computed using
 \begin{equation}
\!\! C^\alpha_\textmd{HBN}(t)\!=\!{\sum_{i_\alpha}\left<\Delta n^{i_\alpha}_\textmd{HB}(t)\Delta n^{i_\alpha}_\textmd{HB}(0)\right>
/
\sum_{i_\alpha} \left<\Delta n^{i_\alpha}_\textmd{HB}(0)^2\right>},
 \end{equation}
 where $n^{i_\alpha}_\textmd{HB}$ is the number of HBs per species $\alpha$ 
(categories \textit{(i)-(iii)} above) and  
$\Delta n^{i_\alpha}_\textmd{HB}(t)\!\!=\!\! n^{i_\alpha}_\textmd{HB}(t)\!-\! \left<n^\alpha_\textmd{HB}\right>$. 
Within the simulation times and  
temperatures investigated, this quantity is fully decorrelated. 
The results for bulk water at 
230, 250 and 270\,K up to 100\,ps  are shown in fig.~\ref{fig:hb-longcorr}.  
Results 
for the three categories of H-bonds 
are
shown in fig.~\ref{fig:hb-corr}. At high temperatures, the HBs in the solvation 
shell persist longer than in bulk: 
The hydrophobic effect dominates.
Solvation shell HBs are also stronger than HBs between water molecules and the carbonyl oxygen;
at higher temperatures the water 
molecules are too fast to form stable HBs with the carbonyl oxygens. 
This agrees well with IR spectroscopy where fast HB dynamics  at few 
hundred femtoseconds has been observed by monitoring the O-D stretch perturbation in the neighbourhood of lipid 
carbonyl groups~\cite{volkov09}, and the observation of moderate HB slowing down in hydrophilic solvation~\cite{sterpone10}.  
For $T\!\!<$265\,K, the water-carbonyl oxygen HBs become longer-lived 
than bulk  and solvation water HBs:
Water-carbonyl binding crossover is located at 
265$\pm$5\,K.  
This is a hydrophobic/hydrophilic crossover and
it has also been reported for water confined in carbon nanotubes 
upon cooling from 295 to 281\,K~\cite{wang08}.

\begin{figure}
\begin{center}
\includegraphics
[width=7.0cm]
{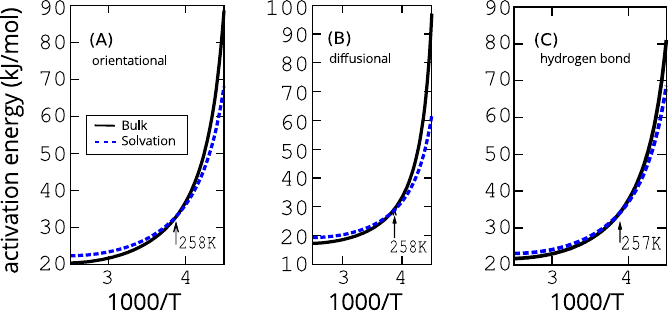} 
\vspace*{-0.2cm}
\caption{
Activation energy crossover 
for A) rotational dynamics (from $C_2(t)$), B)
diffusion (from diffusion coefficient), C) H-bonds
(from single HB lifetime; 
from the correlation 
$C_\textmd{H} (t)\!=\!(\sum_{ij} \left<p_{ij}(t)p_{ij}(0)\right>)/
(\sum _{ij} \left<p_{ij}(0)\right>)$, where $p_{ij}(t)$ 
is the probability that water molecules $i$ and $j$ are H-bonded at time $t$.)}
\label{fig:crossovers}
\end{center}
\end{figure}
 \begin{figure}
\begin{center}
\includegraphics
[width=6.8cm]
{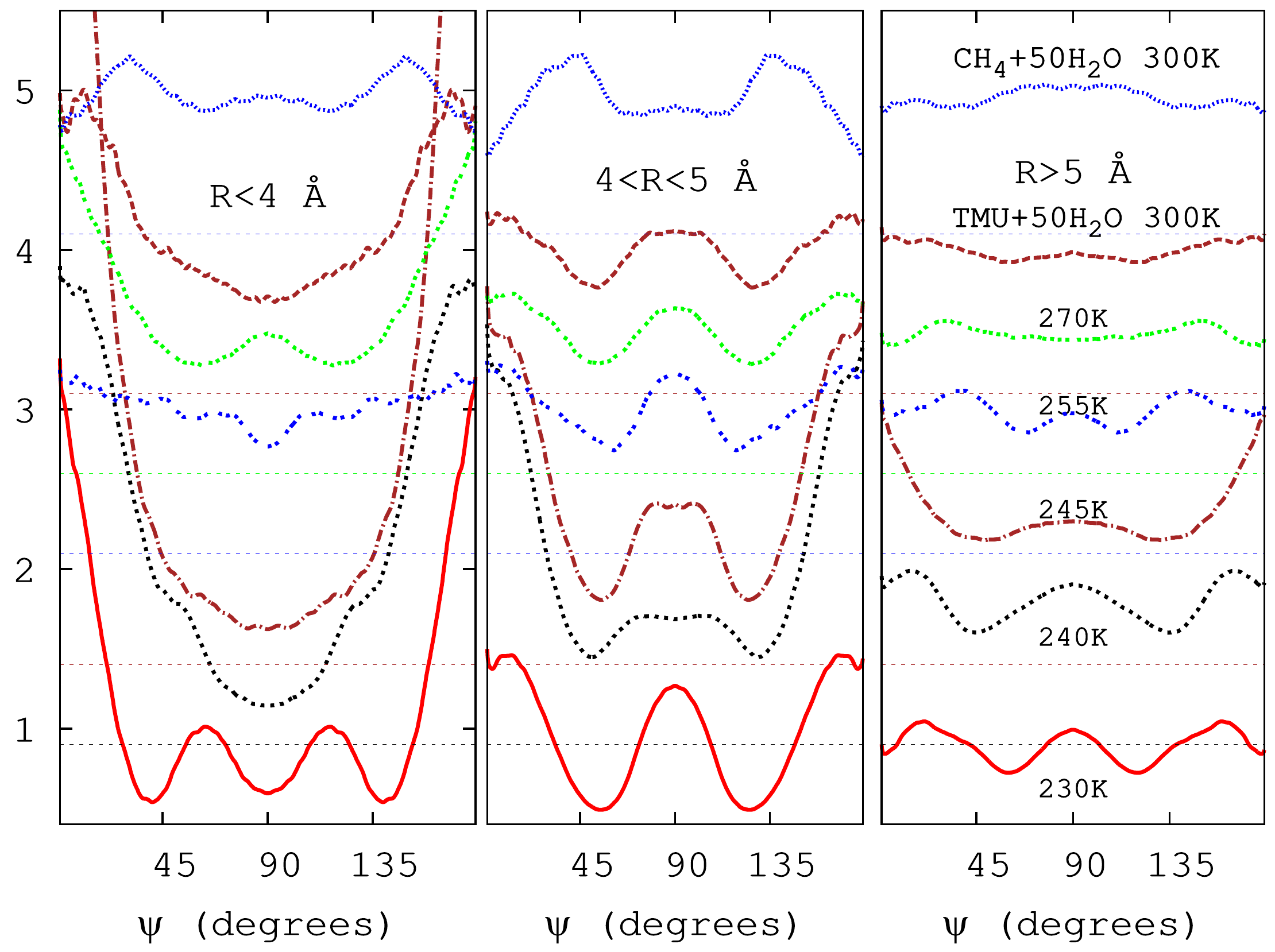} 
\vspace*{-0.2cm}
\caption{Twist angle $\psi$ distribution with  respect to the normal to the surface of the TMU CH$_\textmd{3}$ group. 
Distributions for water surrounding 
a methane (CH$_4$) molecule are also shown for $T$=300\,K. The lines have been shifted 
for clarity.  
There is a  dramatic change in the shape for $T\!<\! 255$\,K. This effect is 
maximal at $T\!\!=\!\!245$\,K where the planes of almost all solvation water molecules are  tangential to 
the hydrophobic surface.}
\label{fig:twist}
\end{center}
\end{figure}

Increasing strength of the carbonyl-water HBs upon cooling 
is in agreement with ref.~\cite{ben-naim13}. 
This trend is crucial for understanding cold denaturation: A consequence of strengthened hydrophilicity 
at low temperatures~\cite{ben-naim13}, rather than purely hydrophobicity as often implied
from the Kauzmann model~\cite{kauzmann59}.  
The strengthening of water-carbonyl HBs upon cooling, evident from the sudden increase in the HB 
strength at 255\,K, is consistent with the finding that a  crossover at 252\,K
for protein solvation water  is due to a large increase of HBs within
the solvation shell~\cite{mazza11,mazza12}.  Such a crossover is not found for bulk water, although 
at 213\,K the relaxation times seem to diverge. Such a temperature cannot 
be studied with current AIMD approach and we therefore term this temperature as "temperature of 
apparent divergence".  The presence of the crossover temperature for solvation water at 256\,K 
pushes the apparent divergence to 190\,K.  The 190\,K crossover for a water monolayer hydrating 
a hydrophobic surface is associated to
the Widom line departing from the liquid-liquid critical point, while the crossover at 252\,K is 
associated to a specific heat maximum that
is consequence of the structural change due to the formation of a macroscopic number of HBs in 
the shell as discussed in ref.~\cite{Bianco2014}. 
Besides understanding water's anomalies and phase 
transitions~\cite{poole92,taschin13,mallamace13,Limmer2014}, its behaviour at supercooled
temperatures has broad practical importance~\cite{Chen2006c,mazza11}. 
For example, unlike in bulk water, in cold denaturation  
water does not crystallize on proteins' surfaces allowing for preservation of cells.
Another example of the importance of understanding water's low temperature behaviour is clathrate formation.
A complete study locating the liquid-liquid 
critical point requires constant pressure 
simulations at various pressures or simulations at various densities and is beyond 
the scope of this work.}
 
The crossover from high activation energy process at higher temperature to low activation energy dynamics 
at lower temperature  is not unique to  OH stretch motion. 
In fig.~\ref{fig:crossovers} we show similar crossover behaviour for the OH rotational dynamics, 
diffusion 
and the single HB life-span. 
Rotational dynamics in fig~\ref{fig:crossovers}A was obtained from 
the correlation function $C_2(t)$ as defined above.
All of the measurements indicate crossovers at temperatures  close
to the experimentally  reported value of 255$\pm$2 K~\cite{qvist08}.
This may be expected since rotational dynamics and diffusivity of water 
are strongly governed by HB dynamics.

\section{Structural properties}

Next, we study the structural changes accompanying the dynamic crossover. 
First, we analyzed the distribution of the angle $\psi$. This angle is 
defined for each water molecule as that between the normal to the surface of a sphere around 
each of the CH$_\textmd{3}$ groups of the TMU molecule (or methane) and the normal 
to the surface formed by the three atoms of the water molecule~\cite{hore08}.  The former normal is on a line 
connecting the carbon atom of the methyl group and the oxygen atom of the water molecule.
This angle, termed twist angle by Hore \textit{et al.}~\cite{hore08}, 
has been used to describe water's rotational motion around proteins~\cite{vogel09}.  
We performed additional 60-200\,ps simulations for the solvation water of a methane molecule 
at 240, 250, 280, 300, 320 and 350\,K. 

The temperature dependence of $\psi$ is shown in fig.~\ref{fig:twist} for the TMU molecule. 
Results for the methane system at 300\,K are also shown as the uppermost curves. 
For $\psi$=0 and  180$^\circ$, the water molecule's plane is tangential to the surface of 
the CH$_\textmd{3}$ group while for $\psi$=90$^\circ$ it is  perpendicular to it. As the rightmost panel 
shows, for the TMU molecule, water molecules beyond 5~{\AA} maintain isotropy at high temperatures. 
But as water molecules get closer 
to the hydrophobic surface, tangential arrangement is preferred, especially at supercooled temperatures.
As the temperature crosses 255\,K, a drastic increase in the fraction of water molecules adopting 
a tangential arrangement appears. At $T\!\!=\!\!245$\,K almost all the water molecules in the 
solvation shell adopt a tangential arrangement. This tangential arrangement of water planes 
together with the tangential dipole moment (discussed below) suggest a sort of carpeting of 
the surface of the hydrophobic  groups by water. 
To test this hypothesis, we computed the connectivity of  the HBs of the water molecules 
that are within 4~{\AA} from the methyl group. We find that for each molecule, 2.0$\pm$0.2 HBs emanate from 
water molecules that are also within 4~{\AA} radius. The other HBs are from  water molecules further away;
dangling bonds are essentially avoided.
These further water molecules are those sequestrating the bulk and the solvation water 
as was found earlier by Hore \textit{et al.}~\cite{hore08}. 
In the case of the solvation water of methane  (about 19 water molecules) 
there are both tangential and sequestrating water molecules in the solvation shell, 
which suggests that this observed behaviour for TMU may be mediated by the hydrophilic carbonyl-water 
interaction.

A Raman study on the OH vibrations of water in alcohols showed an increase in the population of 
water molecules vibrating with lower frequency for solvation water together with the decrease in the population 
of dangling bonds  of the solvation water of hydrophobic interfaces. 
This was interpreted as a signature of increased tetrahedral order for solvation water~\cite{ben-amotz12}. 
Consideration of other order parameters
such as translational order, orientational order and water plane's orientation angle (twist angle) 
reveals different type of ordering for solvation water as compared to bulk water. 
For CH$_4$ and TMU at 300\,K, we find a marginal decrease to values of $Q=0.71$ and 0.65, respectively, 
of  the tetrahedral order of solvation water with respect to bulk water ($Q=0.73$). 
The usual definition of tetrahedral order was applied, i.e.,
$
Q=1-{3\over 8}\sum_{i=1}^{3}\sum_{j=i+1}^{4}\left(\cos \theta_{ikj}+{1\over 3}\right)^2,
$
for 4-coordinated water molecules, where 
$\theta_{ikj}$ is the angle subtended on the central oxygen by the oxygen atoms of water molecules 
$i$ and $j$ which both belong to the nearest neighbour shell of the central water molecule.

\begin{figure}
\begin{center}
\includegraphics
[width=6.5cm]
{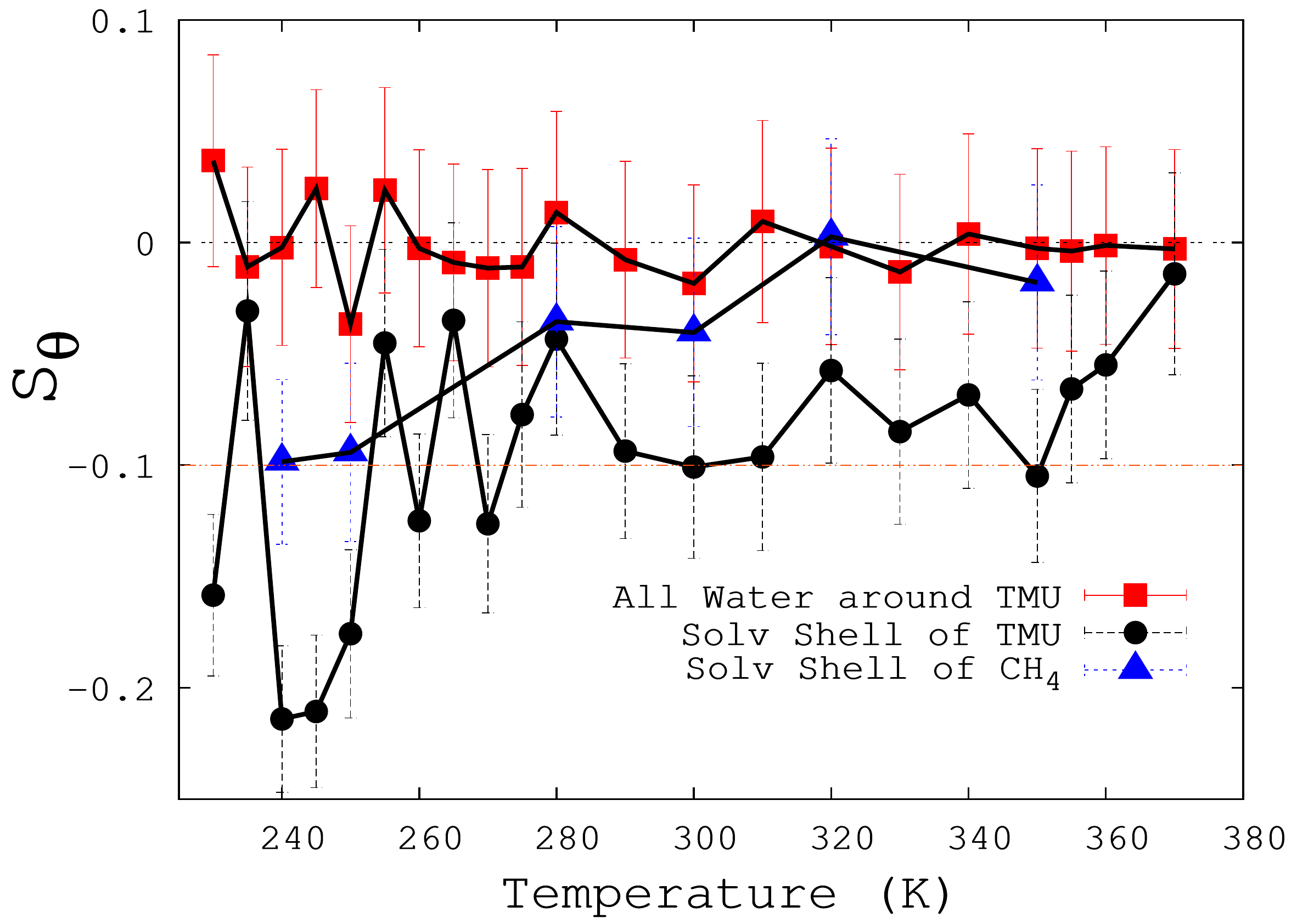} 
\vspace*{-0.2cm}
\caption{The dipole orientational order S$_\theta$ 
for water surrounding the TMU molecule, 
water molecules in the 1$^{\mathrm{st}}$ solvation shell of the  CH$_\textmd{3}$ groups of the TMU, and the solvation shell 
of the methane molecule. The lines are guides to the eye.
}
\label{fig:twist-order}
\end{center}
\end{figure}

Waters' orientational ordering around the TMU methyl groups was characterized by
the 
order parameter
$S_\theta\!=\!{1 \over 2}\left<3\cos^2\theta_i-1\right>_i $,
where $\theta_i$ is the angle the dipole moment of water $i$ makes with the normal to the surface 
of the closest methyl group. For an isotropic system, $S_\theta$=0.
Figure~\ref{fig:twist-order} shows that the average over \textit{all} water 
molecules is $S_\theta \negmedspace \approx$0
and that restricting sampling to the solvation shell gives non-vanishing negative values.
Above 350\,K,  $S_\theta \! \rightarrow \! 0$.
Below 255\,K a significant decrease occurs around 245\,K. 
{A weak but monotonically increasing order around the fully hydrophobic CH$_4$ molecule
(fig.~\ref{fig:twist-order}) is contrasted by the rapid increase in ordering of TMU's solvation water 
as temperature is lowered below the crossover.  The change in the relative strengths of 
the hydrophobic and hydrophilic interactions~\cite{ben-naim13} (solvation free energies) 
around the crossover is the probable reason for the difference.}
These findings, with fig.~\ref{fig:twist},  suggest enhanced structuring for 
TMU solvation water for the range 235-255\,K.
We speculate that this may be related to the stabilized water-carbonyl HBs which are absent 
in the methane solution. We caution, though, that at very low temperatures sampling is limited as 
manifested by the spike below 240\,K. 

\section{Conclusions}

The effect of temperature on water's OH stretch-frequency correlation function was used
to investigate the  proposed dynamic crossover from highly thermally  activated 
vibrational motion of solvation water at higher temperatures to weak temperature dependent 
motion below  256$\pm$4\,K. 
This crossover, attributed to hydrophobic effect, is preceded by one at 265$\pm$5\,K. The latter 
crossover is a hydrophilic one: The weak water-carbonyl HB (with respect to water-water HB) 
at higher temperatures gives way to the stronger water-carbonyl HB. The dynamic 
crossovers are accompanied by structural changes in solvation water, where water 
molecules tend to carpet the surface of the hydrophobes by forming two in-plane HBs and two 
sequestrating HBs that connect solvation water with bulk water. These findings may be related to 
the various structural and thermodynamic transitions reported for a wide range of proteins in solution and 
shed light on the mechanism of cold denaturation of proteins.




\end{document}